# ACCELERATION DETECTION OF LARGE (PROBABLY) PRIME NUMBERS

Dragan Vidakovic[1], Olivera Nikolic[2] and Dusko Parezanovic[3]

[1]Faculty of Business Valjevo (Department Ivanjica), Singidunum University, Belgrade, Serbia
`dragan.vidakovic@open.telekom.rs`
[2]Faculty of Business Valjevo, Singidunum University, Belgrade, Serbia
`onikolic@singidunum.ac.rs`
[3]Club of young mathematicians and computer scientists INFOMAT Ivanjica, Serbia
`infomat@open.telekom.rs`

## ABSTRACT

*In order to avoid unnecessary applications of Miller-Rabin algorithm to the number in question, we resort to trial division by a few initial prime numbers, since such a division take less time.*

*How far we should go with such a division is the that we are trying to answer in this paper?For the theory of the matter is fully resolved. However, that in practice we do not have much use.*

*Therefore, we present a solution that is probably irrelevant to theorists, but it is very useful to people who have spent many nights to produce large (probably) prime numbers using its own software.*

## 1. INTRODUCTION

The basic idea known as Model Driven Security [1, 2], which is the concept of simultaneous system and its security modelling, includes total reliance on protection mechanisms offered by well-known platforms available on the market.

In the paper [3] we suggested that protection by one's own packages, solutions and code should precede large state system protection, otherwise it would not make sense to talk about system protection.

Our intention is to protect data secrecy and integrity by RSA mechanism in which identification of the secret key is equivalent to the problem of prime number factoring. Therefore, our aim is to generate large prime numbers, since their multiplication result, which plays a role in the secret key formation, is a one-way hash function. The question of (probable) primality is solved by Miller-Rabin algorithm [4] and it is a process which takes a lot of time.

What is the optimal number of prime numbers by which we should divide the number in question before Miller-Rabin test application when we would like to find out as soon as possible whether the number is a probable prime or a compound number? Performing division too many times would only be waste of both: time and resources and the shortest period possible for RSA mechanisms parameter generating is a condition of its application in practice.

## 2. WHY DO WE NEED PRIME NUMBERS?

The public key cryptography-PK [4], a major breakthrough in the field of data secrecy and integrity protection, is mostly based on the assurance which has never been mathematically proved that some mathematical problems are difficult to solve. The two of them are particularly prominent and used a lot.

Since we opted for RSA mechanism we will point to one of them. The multiplication of two large prime numbers is a one-way hash function, which means that we can easily get their multiplication result. However, factorization of that multiplication result with the aim of getting the prime factors (factoring), turned out to be very difficult. This problem of identifying private key d in the Public Key cryptography (PK), if we know the public key and if it is the pair (n,e) are two equivalent problems. Certainly, there are many other PK schemes, asymmetrical algorithms, apart from RSA. They are based on the same problem which is difficult to solve in practice if the number of digits is large enough, and by means of these schemes a one-way function with "trap door" is created.

By technological development and progress in the field of algorithms for whole number factorization, the need for larger and larger prime numbers has been demonstrated. This means that their multiplication result will consist of more and more digits. The competition between those who attack unprotected data and those who protect them using RSA mechanism requires creation of the faster operations for dealing with large numbers. The new arithmetic requires more efficient codes for addition, subtraction, multiplication and division of large numbers and what is particularly significant is to solve modular exponentiation in the most efficient way possible.

This makes sense only if special attention is paid to the creation of one's large (probable) prime numbers, since the use of such numbers available on the Internet or in any other way is not in accordance with the very aim of data protection. Since the process of large prime generation requires a lot of time and computer resources, it is of particular interest to us to find a way to avoid the application of the primality testing algorithm to the number as much as possible. We shall solve the problem as it is suggested in [4].

## 3. TASK AND AIM

The relation between the time needed for performing exponentiation ($x^e$ mod n) and the time needed for performing division by one prime appears to be an important parameter in the process of optimization of the time needed for generating probable primes. The question arises by how many first primes the number in question should be divided before the Miller-Rabin algorithm is applied to it.

The trial division takes less time then exponentiation, but it would certainly be wrong to conclude that we should divide the number as long as possible. It is very difficult to determine the real relation between the two, since everything depends on the number we start with and odd numbers we examine so as to generate a probable prime.

Let us assume that we put 59 primes in the table. This means that numbers from 3 to 251 are all stored in the table. We shall observe a 256-bit number in which digit 1 is in the following positions: 0, 1, 2, 3, 4, 5, 6, 7, 10, 50, 100, 127, 255. The aim is to detect the (probably) prime number in which digit 1 is in the following positions: 0, 2, 3, 5, 8, 10, 59, 100, 127, 255. We measured time in seconds as well as the number of test applications to the number in question. They are necessary for detecting the prime number from the initial number, depending on the number of initial primes by which we perform trial divisions. The results are given in the Table 1.

Table 1. Measurement results

| Num. of primes | Time | Num. of passes |
|---|---|---|
| 0 | 295.06 | 24 |
| 1 | 195.98 | 16 |
| 2 | 180.43 | 15 |
| 3 | 180.37 | 15 |
| 4 | 195.87 | 16 |
| 5 | 180.76 | 15 |
| 6 | 180.87 | 15 |
| 7 | 196.14 | 16 |
| 8 | 180.87 | 15 |
| 9 | 180.98 | 15 |
| 10 | 180.98 | 15 |
| 11 | 195.92 | 16 |
| 12 | 181.20 | 15 |
| 19 | 181.86 | 15 |
| 20 | 168.18 | 14 |
| 25 | 168.74 | 14 |
| 30 | 169.23 | 14 |
| 35 | 169.72 | 14 |
| 45 | 170.69 | 14 |
| 59 | 171.97 | 14 |

The question arises whether it is possible to determine the optimal number of trial divisions applications by any analysis of the initial number or any other parameter so that we could minimize the time needed for generating a prime number.

## 4. TIME MEASUREMENT ANALYSIS AND ITS CONNECTION WITH THE PROBLEM SOLUTION

The graph (Figure 1), which shows the relation between time needed for generating a prime (given in seconds) in the function of the number of trial divisions applications to the number we start with, based upon Table 1, suggests that we are obliged to divide the number by at least 3 (since time needed for generating decreases dramatically if we divide numbers by 3 instead of applying Miller-Rabin algorithm to them). The graph shows that it is optimal to divide the number in question by the first 22-23 prime numbers and only then should the Miller-Rabin test be applied to the number we get. The essential question is whether there is a way to determine in advance the optimal number of trial divisions applications knowing only the initial number and the table with the primes.

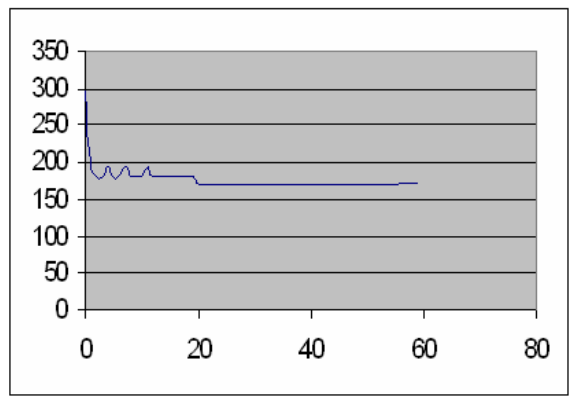

Figure 1. Time dependence of the number of prime in Table 1.

In order to answer the above formulated question we shall measure the time needed for div* of a starting odd number.

Div* is a process of generating an odd number indivisible by any of the first few primes (in this paper we suggest that it should be 59 primes at most). The results are given in Table 2.

Table 2. Measurement results

| Num. primes | Time | Passing |
|---|---|---|
| 1 | 0 | 0 |
| 2 | 0.25 | 0.12 |
| 3 | 0.25 | 0.08 |
| 4 | 0.35 | 0.09 |
| 5 | 0.45 | 0.09 |
| 6 | 0.53 | 0.09 |
| 7 | 0.64 | 0.09 |
| 8 | 0.73 | 0.09 |
| 9 | 0.84 | 0.09 |
| 10 | 0.85 | 0.08 |
| 11 | 0.94 | 0.08 |
| 12 | 1.05 | 0.09 |
| 13 | 1.17 | 0.09 |
| 14 | 1.26 | 0.09 |
| 15 | 1.36 | 0.09 |
| 16 | 1.46 | 0.09 |
| 17 | 1.56 | 0.09 |
| 18 | 1.64 | 0.09 |
| 19 | 1.75 | 0.09 |
| 20 | 3.6 | 0.18 |
| 21 | 3.7 | 0.18 |
| 22 | 3.82 | 0.17 |
| 25 | 4.17 | 0.17 |
| 30 | 4.7 | 0.16 |
| 35 | 5.18 | 0.15 |
| 45 | 6.08 | 0.14 |
| 59 | 7.42 | 0.12 |

The analysis of Table 2 can be presented in the best way in the form of a graph (Figure 2). We put the number of prime numbers by which we divide the initial number (as well as the other numbers) on the coordinate axis x whereas we put the average value of the time needed for division on the coordinate axis y (the quotient of time needed for division and number of the first prime numbers by which we divide the initial number in seconds).

The graph in fact represents the curve of the usefulness of the operation of div*. When we say usefulness we imply the longest period needed for the division of a number which therefore means the shortest period needed for the Miller-Rabin test and that is in fact our aim.

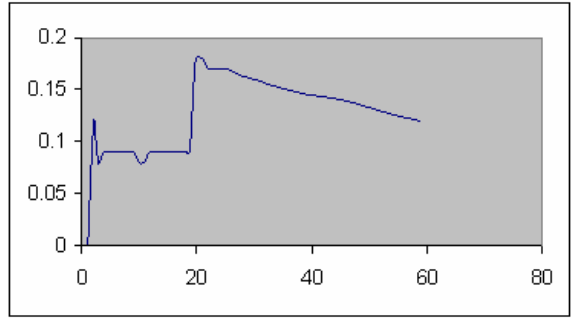

Figure 2. Curve usefulness

It is obvious that the optimal bound to which we perform trial division is actually the point of absolute maximum of the function of div* usefulness. If we look at it, we shall notice that it is very useful (local maximum) when we divide by 3 and 5, and 3, 5, 7 which means that in this case division by first three numbers is obligatory. The next bound is the point of the new local (in this case it is absolute as well) maximum of usefulness, and that is the point 22-23. In the concrete case it means that the division by the first 22-23 prime numbers combined with the primality testing leads to the fastest detection of a probably prime.

The question arises what is the optimal bound is, i.e. by how many first primes it would be optimal to divide a number before we apply primality test. If we develop the original idea from [4] *("(optimal trial division bound B) Let E denote the time for full k-bit modular exponentiation, and let D denote the time required for ruling out small primes as divisor of a k-bit integer.* (*The values E and D depend on the particular implementation of long integer arithmetic) Then the trial division bound B that minimizes the expected running time of algorithm for generating k-bit prime is roughly B=E/D. A more accurate estimate of the optimum choice for B can be obtaind experimentaly. The odd primes up to B can be precomputed and stored in a table. If memory is scarce, a value of B that is smaller than the optimum value must be used*"), not so strict a bound can be number B=E/D, where E denotes the period of time needed for the application of the Miller-Rabin test to the prime number (the time measured on PC-486 is 15.92''), without divisions by small primes, whereas D denotes maximal usefulness (from Table 2 into Figure 2, it can be seen that the time is 0.18''). Consequently B=15.88/0.18=88. This means that the div* by all prime numbers smaller then 88 is desirable. Since there are 24 such numbers it means that each number we get should be divided by the first 24 primes and only then should the number prepared in that way be tested for primality. It is clear that this result matches the experimental results and it can certainly be used as an orientation bound when we have to estimate how to make the process of generating large probably primes faster.

If we start with the closer number i.e. if digit 1 is in the positions 0, 2, 4, 8, 10, 50, 100, 127, 255 the figure which represents the time needed for reaching the prime number looks like this (Figure 3.)

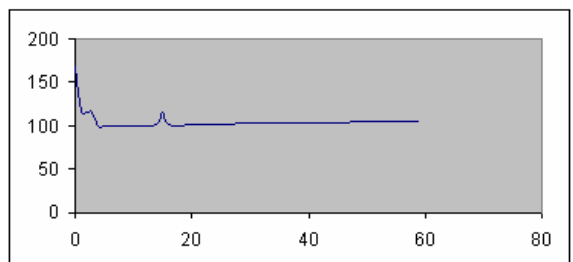

Figure 3. Graph of the previous example

The figure of the div* usefulness (Figure 4) looks like this in this case:

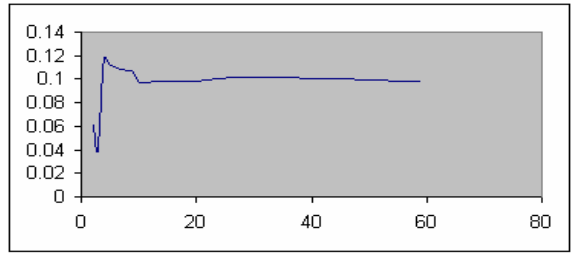

Figure 4. Curve usefulness

The analysis of the graph shows that the most useful thing to do is to divide the initial number by numbers 3, 5, 7 and 11 because that is where the maximum of div* of usefulness is (at point 4 the maximum value is 0.117). On the other hand, according to the estimate E/D=15.92/0.117=136 it is expected that the best thing to do is to divide the number by the prime smaller then 136, and it is 131. According to the graph it is clear again that it is an acceptable estimate and that a quotidian E/D can be taken for the above mentioned bound.

It is interesting to analyse the function of div* usefulness using as an the process of generating a 128-bit number. Let the initial number contain the digit 1 in the following positions: 0, 1, 2, 8, 10, 50, 100, 127. Let us measure the time needed for generating the prime number with digit 1 in the following positions: 0, 1, 2, 3, 4, 6, 8, 10, 50, 100, 127. The time of div* usefulness is given in Table 3.

Table 3. Measurement usefulness

| Num. primes | Time |
|---|---|
| 1 | 0 |
| 2 | 0.085 |
| 3 | 0.05 |
| 4 | 0.055 |
| 10 | 0.055 |
| 25 | 0.062 |
| 35 | 0.063 |
| 45 | 0.064 |
| 59 | 0.063 |

The graph of the function of div* usefulness is given in Figure 5.

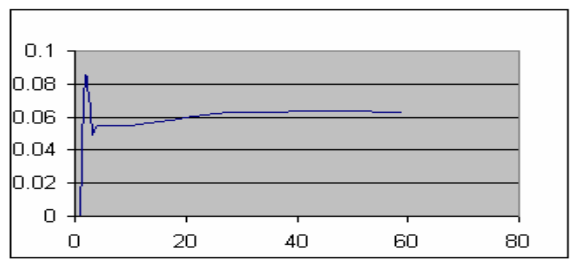

Figure 5. Curve usefulness

The task is to determine the optimal number of trial divisions applications needed for generating a probably prime number as quick as possible. The real hypothesis is that is the point at which the function of the div* usefulness reaches absolute maximum, using the numbers from the given table.

Table 3 or the graph shows that the function is maximal at point 2 and it is value is 0.085. Consequently, it would be best to divide the initial number, and the numbers which appear before we reach the prime by the first two primes (i.e. numbers 3 and 5). Let us check our hypothesis now, let us measure the time needed for reaching the given prime number if we start from the above given number. In this process we perform trial division from 1 to 59 first prime numbers.

Table 4. Measurement results

| Num. primes | Time | Passing |
|---|---|---|
| 0 | 76.5 | 45 |
| 1 | 51.25 | 30 |
| 2 | 49.26 | 29 |
| 3 | 49.28 | 29 |
| 10 | 49.46 | 29 |
| 20 | 49.75 | 29 |
| 30 | 50.08 | 29 |
| 40 | 50.37 | 29 |
| 59 | 51 | 29 |

The graph which represents the function of total times (in seconds) in the function of the number of the first primes by which we divide the number in question (Figure 6) trying to identify the prime number looks like this:

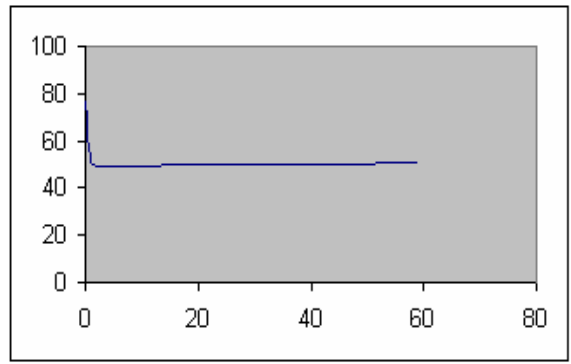

Figure 6. Curve usefulness

We really did expect the best result to be at the point 2, i.e. it is optimal to divide by numbers 3 and 5 (the first two primes). We checked experimentally the hypothesis that the (absolute) maximum of the function of div* usefulness is a bound and that we should perform trial divisions until we reach it. It is easy to solve the problem by means of a program; what we need to do is subject a chosen odd number to the process of division and then find the average value. After that we should take the point at which that value is the highest. It will be the bound which denotes the number of the first primes by which we divide all numbers until we reach a probably prime number, before we apply primality test to the number.

Another improvement could have been performed by taking odd numbers which are members of specific arithmetic progressions which contain endless series of prime numbers, but we did not opt for that in the paper.

## 5. CONCLUSION

The important stage in the process of changing of the basic paradigm known as MDS, by replacing or supplementing available solutions for security mechanisms by one's own solutions, is the process of generating large primes. We believe that we solved the task suggesting some ideas in this paper and reached an important aim, this being the saving of all computer resources needed for that process. We believe that by means of the procedure we presented, we significantly reduced unnecessary use of primality testing procedure and thus made an important step in the process of creating of one's own hardware-software protection mechanism.